\pdfoutput=1

\documentclass[11pt]{article}
\usepackage{footnotebackref}
\usepackage{acl}
\usepackage{booktabs}
\usepackage{times}
\usepackage{latexsym}
\usepackage{multirow}%
\usepackage[T1]{fontenc}

\usepackage{setspace}
\usepackage{changepage} 
\usepackage[breakable]{tcolorbox}
\usepackage{float}

\usepackage[utf8]{inputenc}
\usepackage{graphicx}
\usepackage{microtype}
\usepackage{xspace}

\newcommand{\engx}{{eng--X}\xspace}

\newcommand{\etox}{\textsc{etox}\xspace}
\newcommand{\detoxify}{\textsc{detoxify}\xspace}

\newcommand{\asretox}{\textsc{ASR-ETOX}\xspace}

\newcommand{\whisperlarge}{\textsc{Whisper-Large-v2}\xspace}

\newcommand{\commonvoice}{\textsc{CommonVoice}\xspace}
\newcommand{\expressivemined}{\textsc{SeamlessAlignExpressive}\xspace}
\newcommand{\speechmined}{\textsc{SeamlessAlign}\xspace}

\providecommand\marta[1]{[\textcolor{magenta}{Marta: {#1}}]}

\newif\ifdraft
\draftfalse

\usepackage{newfloat}
\usepackage{listings}
\usepackage{booktabs}
\usepackage{comment}
\usepackage{hyperref}
\usepackage{url}
\usepackage{algorithm}
\usepackage{algorithmic}
\usepackage{amsmath}
\usepackage{multirow}
\usepackage{graphicx}
\usepackage{xcolor}
\usepackage{soul}
\usepackage{pgf}
\usepackage{color, colortbl}
\title{MuTox: Universal MUltilingual Audio-based TOXicity \\ Dataset and Zero-shot Detector}

\author{Marta R. Costa-jussà, Mariano Coria Meglioli, Pierre Andrews, \\ \textbf{David Dale, Prangthip Hansanti, Elahe Kalbassi,}\\ \textbf{Alex Mourachko, Christophe Ropers, Carleigh Wood}\\
FAIR, Meta \\
\texttt{\{costajussa,mfcoria,mortimer,daviddale,prangthiphansanti,}\\
\texttt{ekalbassi,alexmourachko,chrisropers,carleighwood\}@meta.com}}

\begin{document}
\maketitle

\begin{abstract}
Research in toxicity detection in natural language processing for the speech modality (audio-based) is quite limited, particularly for languages other than English. To address these limitations and lay the groundwork for truly multilingual audio-based toxicity detection, we introduce MuTox, the first highly multilingual audio-based dataset with toxicity labels which covers 14 different linguistic families.
The dataset comprises 20,000 audio utterances for English and Spanish, and 4,000 for the other 28 languages. To demonstrate the quality of this dataset, we trained the MuTox audio-based toxicity classifier, which enables zero-shot toxicity detection across a wide range of languages.
This classifier performs on par with existing text-based trainable classifiers, while expanding the language coverage more than tenfold. When compared to a wordlist-based classifier that covers a similar number of languages, MuTox improves F1-Score by an average of 100\%. This significant improvement underscores the potential of MuTox in advancing the field of audio-based toxicity detection.

\end{abstract}

\textit{{\color{blue}Warning: This article includes examples of language that can be considered offensive or upsetting.}}

\section{Introduction}

\ifdraft
    \marta{TODO list
    \begin{itemize}
        \item discussion of biases in data card + other details in the cards
        \item results in unbiased task
        \item add examples in appendices
    \end{itemize}
}

\else

Text toxicity detection has been largely explored for different tasks in Natural Language Processing (NLP) \cite{jigsaw-dataset}. Wordlist-based toxicity classifiers---e.g., \etox \cite{costajussa2023toxicity}--- scale well to a large number of languages \cite{nllb2022} and context-based classifiers are able to detect beyond lexical toxicity with tools such as \detoxify\footnote{\label{detoxify}https://github.com/unitaryai/detoxify}. 

When exploring audio-based toxicity detection, there are either cascaded systems which extend text toxicity detection with speech recognition \cite{communication2023seamlessm4tmassively}; or end-to-end audio-based toxicity classification \cite{Ghosh2021DeToxyAL} which provides an English dataset together with end-to-end toxicity detection results. This work shows that gains of English text-less audio-based classifiers over text-based classifiers are specially relevant when applied to out-of-domain, coherently with the previous study on a non-disclosed dataset \cite{DBLP:conf/eusipco/YousefiE21}.  

In this paper, we go far beyond existing research in audio-based toxicity detection by providing the first highly multilingual audio-based toxicity annotated dataset (MuTox dataset, 30 languages, see Table \ref{table:language_list} in appendix \ref{app:lang}) together with the first text-less massive multilingual metric (MuTox classifier, 100+ languages). Note that multilinguality for audio-based toxicity detection becomes even more crucial for the task of added toxicity in the context of multimodal and multilingual translation, where the case of adding or deleting toxicity may be considered as a critical error \cite{communication2023seamlessm4tmassively}.

In particular, the main contributions of this paper are: providing guidelines for audio-based toxicity annotation (section \ref{sec:guidelines}); releasing the first highly multilingual audio-based toxicity dataset and benchmark with human annotations for 30 languages (section \ref{sec:dataset}); analyzing the performance of text-based classifiers when applied to audio-based toxicity detection (section \ref{sec:analysis}); proposing MuTox, a massively multilingual audio-based toxicity classifier (section \ref{sec:speechclass}). 
Our results can be summarized: 

\begin{itemize}
    \item When compared to the strongest performing systems, which are composed of speech recognition plus trainable text-based toxicity detectors, MuTox performs on par,  while offering more than 10 times the language coverage \footnote{We want to clarify that while MuTox dataset evaluates 30 languages, the MuTox classifier relies on SONAR \cite{Duquenne:2023:sonarexp_arxiv} embeddings. As of March 2024, SONAR encoders are available for 57 languages in speech and 200 in text. Thanks to this architecture, by design, the MuTox classifier provides zero-shot toxicity detection for all of them. This is more than 8 times language coverage in speech and more than 25 times in text.}. 
\item When compared to systems with the highest coverage, which are composed of speech recognition plus wordlists toxicity detectors, MuTox improves F1-Score by an average of 100\%. 
\end{itemize}

These results highlight the effectiveness of MuTox in multilingual audio-based toxicity detection, demonstrating its potential to significantly advance the field.



\section{Background}
\label{sec:background}

\paragraph{Text-based toxicity classifiers.} In this work, we use \etox \cite{costajussa2023toxicity} and \detoxify as primary text-based toxicity detectors. \etox is chosen for offering the widest language coverage in toxicity detection. \detoxify is chosen for being one of the available tools with the highest performance on several JigSaw benchmarks with a single model \cite{detoxify-justification}.

\etox is an open-source\footnote{\label{etox}\url{https://github.com/facebookresearch/seamless_communication/tree/mutox/src/seamless_communication/cli/toxicity/mutox}} wordlist-based classifier covering 200 languages. While this metric has several advantages - namely, it covers a massive number of languages and is highly transparent - it has other limitations such as only detecting lexical toxicity and not distinguishing polysemous words which may be toxic in some contexts and not others. 

\detoxify\footref{detoxify} is a text-based toxicity classifier trained mainly in JigSaw \cite{jigsaw-dataset} which is freely available in 7 languages (English, French, Italian, Portuguese, Russian, Spanish, Turkish).

\paragraph{Datasets} 
A well-known resource for textual toxicity detection is the JigSaw dataset \cite{jigsaw-dataset}, which consists of a large number of Wikipedia comments which have been labeled by human raters for toxic meaning. This dataset has been used for several Kaggle competitions covering a broad range of tasks, from detecting types of toxicity to analyzing bias in toxicity detection systems. There are also other related datasets e.g. \cite{rottger-etal-2022-multilingual}.

There are extremely few speech datasets with toxicity labels. 
{Recently, ADIMA covers multilingual profanity detection audio dataset in 10 Indic languages \cite{9746718}. }
Relatedly, Detoxy \cite{Ghosh2021DeToxyAL} estimated the amount of toxicity for several English spoken datasets using text-based classifiers. In this work, we follow a similar approach for pre-selecting data for annotation to maximize chances of the annotators confirming toxicity. We used two spoken datasets which cover diverse domains and are highly multilingual: \commonvoice \cite{ardila-etal-2020-common} and \speechmined \cite{communication2023seamless}. 

\commonvoice is a massively-multilingual collection of transcribed speech intended for speech technology research and development. \speechmined is an automatically collected pairs of natural speech from raw web corpora through parallel data mining following methodology described in \cite{communication2023seamlessm4tmassively}. Additionally, for English and Spanish, we use \expressivemined which extends the process used for \speechmined to create a large collection of
multilingual speech/speech and speech/text pairs, aligned not only in meaning but also expressivity. Text and audio sources are
identical to \speechmined. The modified mining algorithm is described in section 4 from \cite{communication2023seamless}.

\paragraph{Experimental Task.} Toxicity detection in natural language processing is the task of assigning a toxicity label to a speech or text utterance. 

\section{Annotation Guidelines}
\label{sec:guidelines}

This section reports the detailed guidelines that we provide to annotators to detect toxic content in audio speech at the level of single utterances. This toxicity could be due to aspects of lexical semantics or of perlocutionary effects.  For this we provide annotators with the following definitions:

\begin{itemize}
    \item \textbf{Utterance} refers to a unit of audio speech that is comparable to what a sentence is for writing. 

\item \textbf{Lexical semantics} refers to meaning clearly attached to a particular word or phrase, as opposed to meaning that can be conveyed by other aspects of audio speech than words, such as by tone of voice. 

\item \textbf{Perlocutionary} refers to the effect that an utterance has on the interlocutor or listener (as opposed to the locutionary or illocutionary aspects of the same utterance). For example, if  Interlocutor A says "it would be a shame if something happened to your family" to Interlocutor B, the utterance has a locutionary aspect (its literal meaning; i.e. something could happen to your family, and that would be bad), an illocutionary aspect (the thinly veiled threat against Interlocutor B's family; i.e. if you don't do as I say, something bad will happen to your family), and a perlocutionary effect (the fear it causes to Interlocutor B).

\end{itemize}

\paragraph{What qualifies as toxicity for this task?} The notion of toxicity can be elusive because of its inherent subjectivity; i.e. it greatly depends on the lived experience of the person who is on the receiving end of an utterance. For the task at hand, we arbitrarily considered the following cases as toxic:
\begin{itemize}
    \item \textbf{Profanities} include slurs and language that is regarded as obscene, repulsive, or excessively vulgar, as well as scatological. Examples of profanities in English include words such as \textit{shit}, \textit{asshole}, \textit{fucking}, etc.
    \item \textbf{Hate speech} is language that is used to demean, disparage, belittle, or insult groups of people. Hate speech in English includes words and expressions such as \textit{women are sluts}, \textit{men are trash}, \textit{wetbacks}, \textit{towelheads}, etc.
    \item \textbf{Pornographic language} is language that refers to sexual acts or refers in a vulgar way to body parts typically associated with sexuality. Examples of pornographic language include words or expressions such as \textit{suck my dick}, \textit{cumshot}, etc.
    \item \textbf{Physical violence or bullying language} is language that is used to bully, threaten, silence individuals. Examples of such language include words or expression such as \textit{son of a bitch}, \textit{shut the fuck up}, \textit{retard}, etc.
\end{itemize}

\paragraph{What does not qualify as toxicity for this task?} Annotators were instructed to refrain from labeling as toxic the following types of language:
\begin{itemize}
    \item \textbf{Common innocuous slang} is language that is particularly colloquial but isn't offensive; e.g., using the word \textit{cops} to refer to police officers.
    \item \textbf{Appearance of hate} shows up as language that conveys hate but isn't directed to other human beings; e.g., \textit{I hate this movie!}
    \item \textbf{Appearance of obscenity} shows up as language that sounds similar to profanities or pornographic language but isn't directed towards people; e.g., \textit{school sucks!}
\end{itemize}

To complete the task, annotators were asked to answer the below questions:  
\begin{enumerate}
    \item Does the utterance contain toxicity?
    \item If you answered positively to Question 1, is the toxicity related to:
    \begin{enumerate}
        \item Specific words or phrases? If so, please specify the word or phrase.
        \item Specific perlocutionary effects? If so, please specify which effect among the following options:
        \begin{itemize}
            \item  aggressive raising of the voice
            \item aggressive tone 
            \item (veiled) threat
        \end{itemize}
    \end{enumerate}
\end{enumerate}

To help annotators reach a faster determination as to which words or phrases could qualify as toxic in this exercise, we also pointed them to the publicly available Toxicity-200\footnote{https://tinyurl.com/NLLB200TWL} word list repository.


\section{MuTox DataSet Description}
\label{sec:dataset}

MuTox is composed of 30 languages. Two languages, English and Spanish, contain a larger amount of annotated data (20k utterances/sentences each language), while the rest of languages (see section \ref{sec:19lang}), contain a smaller amount of annotated data (4k utterances/sentences each language).

\subsection{Annotation for English and Spanish}
\label{sec:datasetengspa}

\paragraph{Preliminary selection.} We used speech transcriptions of datasets and a text toxicity classifier.  
We use mainly highly multilingual datasets: \commonvoice\footnote{https://commonvoice.mozilla.org/} and \speechmined \cite{communication2023seamless}, in addition to the English/Spanish \expressivemined dataset, presented in Section \ref{sec:background}.
In these datasets we use the text-based toxicity classifier \detoxify, presented in section \ref{sec:background}, 
to detect toxicity in the transcribed text.  

We screen audio files by length between 2 and 8 seconds for reasons that relate to both semantic and cognitive loads. On the one hand, annotators reported that it was particularly difficult to assert the meaning of very short utterances (under 2 seconds). On the other hand, longer utterances (over 8 seconds) can have too much cognitive load or too much information to annotate.  For pre-selecting toxicity samples, we perform sampling across \detoxify toxicity categories.
  For pre-selecting clean samples, we use cases where all toxicity scores fall below 0.5.

\paragraph{Data statistics.} Annotation results are reported in Table \ref{tab:datastatistics}. The percentages of confirmed toxic samples are 16 \% and 19\% for English and Spanish, respectively. Figure \ref{fig:categories} reports the types of toxicity for each language using annotated categories.
The proportion of toxicity increases with the toxicity quantile (see Figure \ref{fig:quantiles} in appendix \ref{app:quantiles}) . The correlation coefficient between the toxicity in the annotation with \detoxify threshold is 0.6.

\begin{table}[h!]
    \centering
    \small
    \begin{tabular}{lllll}
    \toprule
&\multicolumn{2}{c}{English} & \multicolumn{2}{c}{Spanish}\\
& utterances  & hours & utterances & hours \\\midrule 
Total & 20K & 21 & 20K & 22.2 \\
Cannot say & 547 & 0.52 & 391 & 0.41 \\
No-Toxicity & 16210 & 17.1 & 15709 & 17.5 \\
Toxicity & 3243 & 3.42 & 3900 & 4.24 \\
\bottomrule
    \end{tabular}
    \caption{Results on toxicity annotated results for English and Spanish.}
    \label{tab:datastatistics}
\end{table}

\begin{figure}[ht!]
\center
    \includegraphics[width=7.9cm]{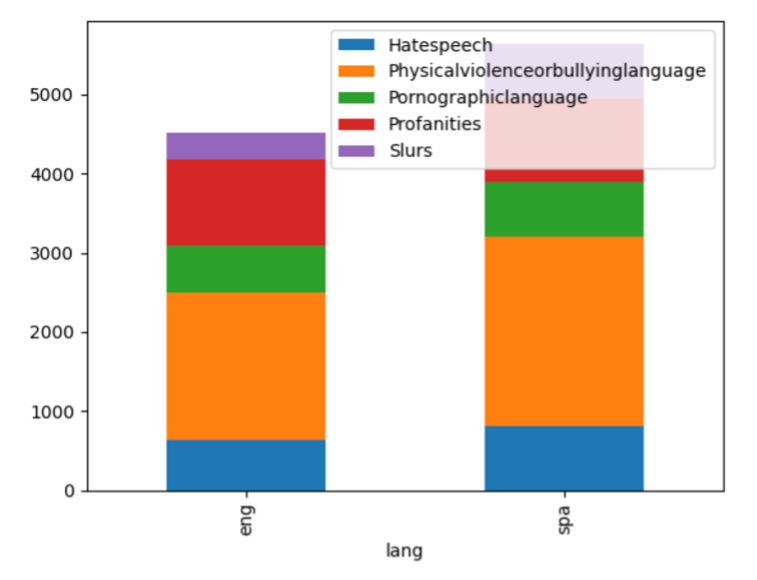}
    \caption{Amount of toxicity per toxic category proposed in this paper.}
    \label{fig:categories}
\end{figure}

\paragraph{Dataset splits.} The final MuTox dataset includes the 20k annotated English and Spanish corpora from Table \ref{tab:datastatistics}. We supplement this with the English \commonvoice annotated data released in Detoxy \cite{Ghosh2021DeToxyAL}. We split the data in subsets of training, dev, devtest and 3k test with stratified samples of types of toxicity and corpora. Detailed statistics are given in Table \ref{tab:datasetspeech}.

\begin{table}[ht!]
\centering
\scriptsize
\begin{tabular}{llllrr}
\toprule
Subset & Language & Modality & Dataset & Size & Toxicity \\ \midrule
\multirow{4}{*}{Train} & \multirow{3}{*}{Eng} & \multirow{2}{*}{Speech} & MuTox  & 13617 & 2270  \\
& & & Detoxy   & 9818 & 2455  \\ 
& & Text & Jigsaw & 21924 & 2928  \\ 
& Spa &  \multirow{2}{*}{Speech} &  \multirow{2}{*}{MuTox} & 13726 & 2730  \\
& HP &  &  & 1212-1498 & 39-363  \\
\midrule
\multirow{3}{*}{Devtest} & Eng & \multirow{3}{*}{Speech}  & \multirow{3}{*}{MuTox} &  {973} & {162}  \\ 
& Spa &  &  & {981} & {195}  \\ 
& HP & &  & 203-250& 7-60 \\
\midrule
\multirow{3}{*}{Devtest} & Eng & \multirow{3}{*}{Speech} & \multirow{3}{*}{MuTox} & 1945 & 324  \\ 
& Spa &  & & {1960} & {390}  \\  
& HP &  &  & 606-749 & 20-182   \\
\midrule
\multirow{3}{*}{Test} & Eng &\multirow{3}{*}{Speech}  
&  \multirow{3}{*}{MuTox} & 2918 & 486  \\
& Spa & &  & 2918 & 486  \\ 
& HP &  &  & 1213-1499 &  38-362 \\
\bottomrule
\end{tabular}
\caption{Audio speech utterances specified by dataset subset\label{tab:datasetspeech}. HP languages footnote. MuTox is our new labelled data that has been annotated in this work. Additionally, we used data from Detoxy \cite{Ghosh2021DeToxyAL} and JigSaw \cite{jigsaw-dataset}. }
\end{table}

\subsection{Annotation for 28 additional languages}
\label{sec:19lang}

\paragraph{Preliminary selection.}  Based on annotated data in English and Spanish (section \ref{sec:datasetengspa}), we devise new criteria to select 4k sentences across 28 extra languages which we consider high-priority (HP) for the context of this work and related previous projects \cite{communication2023seamlessm4tmassively}. The purpose of this extension is to form a highly multilingual benchmark for audio-based toxicity classification. 

\textit{Languages}. Modern Standard Arabic, Bengali, Bulgarian, Catalan,  Czech, Danish, Dutch, Estonian, Finish, French, German, Greek, Hebrew, Hindi, Hungarian, Indonesian, Italian, Mandarin Chinese, Persian, Polish, Portuguese, Russian,  Slovak, Swahili, Tagalog, Turkish, Urdu, Vietnamese (see table \ref{table:language_list}).


\textit{Classifiers}. We use a text toxicity classifier that covers all these languages (\etox) in combination with the English toxicity classifier used in section \ref{sec:datasetengspa} (\detoxify). 

\textit{Datasets}. We use \speechmined{} which has parallel data \engx for all languages of interest.

\textit{Methodology}. We aim at finding 2,500 sentences with the n-largest toxicity scores, and randomly sample the dataset to complete the 4k set. Motivated by results from section \ref{sec:datasetengspa}, sentences are selected as follows:

We include all the samples positive per \etox and \detoxify scores higher than 0.8 (i.e. intersection). This results in a small number of samples, ranging from more than 2k (pol) to 0 (cmn,jpn).
Then, we include \etox detections (ignoring the \detoxify threshold) but allow only for a maximum of 200 sentences for each \etox token and add a maximum of 1,000 \etox sentences. We set these maxima so as to ensure toxicity diversity in the selection.
Finally, we include samples detected with \detoxify:
we use \detoxify in Italian, Portuguese, Turkish, Russian and French (which are supported by the tool). For the rest of non-supported languages, we resort to English \detoxify in the parallel text. The effectiveness of this method is supported by the following evidence: when using \detoxify in English with a threshold of 0.80 on the corpus from section \ref{sec:datasetengspa}, we observe a total toxicity of 0.25 in Spanish.

\begin{figure}[ht!]
\center
    \includegraphics[width=7.9cm]{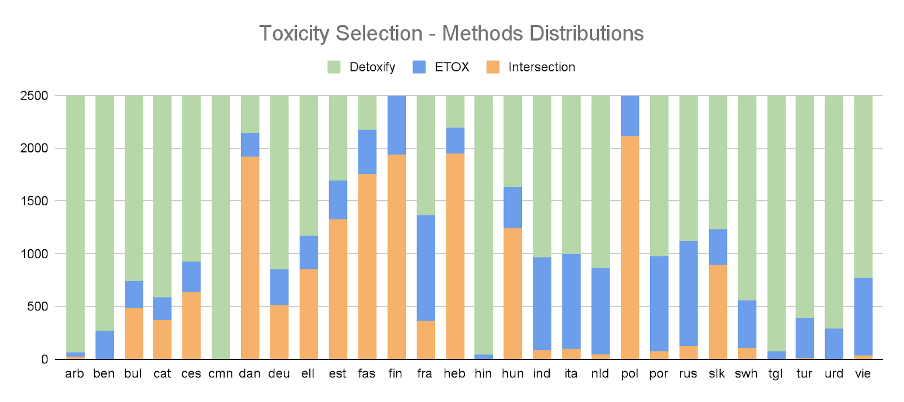}
    \caption{Toxicity selection distribution per language (x-axis) and for each method: \detoxify, \etox and their intersection.}
    \label{fig:19selection}
\end{figure}

\paragraph{Dataset statistics and splits.} Figure \ref{fig:19selection} shows the results of this selection,  grouped by the methodology used (1 - intersection, 2 - \etox, 3 - \detoxify). Table \ref{tab:datasetspeech} shows the dataset splits. The final dataset of 28 languages includes 83 hours.

\section{MuTox audio-based toxicity Classifier}
\label{sec:speechclass}

\paragraph{Methodology: MuTox Classifier.} 
 We feed our toxicity classifier, MuTox, with both audio-based and text-based toxicity-labeled data. The audio-based toxicity classifier depicted on Figure \ref{fig:classifier} follows a simple architecture consisting of an encoder, turning input text or audio speech into a fixed-size representation vector, and a binary classifier composed of three simple feed-forward layers.

\begin{figure}[ht!]
\center
    \includegraphics[width=.9\linewidth]{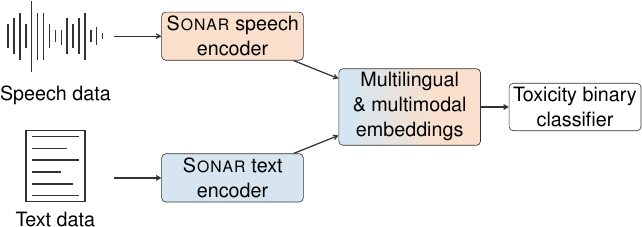}
    \caption{MuTox toxicity classifier diagram \label{fig:classifier}}   
\end{figure}

\paragraph{MuTox Implementation.} We use multimodal and multilingual SONAR encoders from \cite{Duquenne:2023:sonarexp_arxiv}. This choice is motivated by the broad language coverage at the time of this study (English, Spanish, and HP languages) and the zero-shot capabilities of SONAR: once trained on a set of languages, the classifier head can be used on top of any compatible SONAR encoder handling another language. For the classifier, we use variable input sizes for the 3 feedforward layers (1024, 512, and 128). We train with a Binary Cross Entropy loss with logits and Adam optimizer with an initial learning rate of 0.001. In order to compare zero-shot (ZS) vs supervised performance, we train a first classifier with English and Spanish training data only and test on HP languages. We then train a second classifier on all training data available and report results on the same HP languages. The number of parameters of our models is around 600k. Results are listed in Table \ref{tab:toxicity:aucresults}. Throughout our experiments, we use \whisperlarge to generate transcriptions.

\section{Experiments and Results}
\label{sec:analysis}

\paragraph{Experimental Framework.} We evaluate MuTox (text and speech) on Devtest and Test from Table \ref{tab:datasetspeech}. We compare the performance against ASR-\detoxify (hereinafter, \detoxify for simplicity) using the available tool\footref{detoxify} and ASR-\etox (hereinafter, \etox) using the available tool\footref{etox}. Speech recognition is done with \whisperlarge. We report Area Under the Curve (AUC), Precision, Recall and F1-Score across toxicity detection.

\paragraph{Correlation across classifiers} We computed Pearson correlation across classifiers (see results in Table \ref{tab:correlations}). 
We observe that correlation is higher between MuTox and \detoxify than MuTox and \etox or even \detoxify and \etox. In future work, we would like to do a manual analysis of the differences across classifiers.

\begin{table}[h!]
    \centering
    \scriptsize
    \begin{tabular}{lllll}
    \toprule
    & \etox & \detoxify & MuTox & ASR-MuTox \\ \midrule
    \etox &	1	&0.26&	0.06&	0.12 \\
\detoxify &	0.26	& 1&	0.46&	0.49\\
MuTox&	0.06&	0.46&	1	&0.79 \\
ASR-MuTox&	0.12&	0.49&	0.79&	1\\
\bottomrule
    \end{tabular}
    \caption{Pearson correlation across toxicity classifiers.}
    \label{tab:correlations}
\end{table}

\paragraph{Improvements of MuTox compared to one of the Strongest Quality Text-based Toxicity Classifier} 
Table \ref{tab:toxicity:aucresults} compares MuTox and Detoxify (for available languages) in terms of AUC. When comparing the 7 languages covered by \detoxify, MuTox trained on all languages shows on par results with \detoxify. MuTox however, scales to 10 times more languages than \detoxify.

\paragraph{Improvements of MuTox compared to the Text-based Toxicity Classifier with the Largest Coverage} Table \ref{tab:toxicity:precrecallresults} compares MuTox and \etox in terms of recall at fixed precision. ASR-MuTox with a fixed precision of $max (\asretox, 0.3)$ (meaning 0.3 average precision)\footnote{Except for swh and
cat where precision is of $max (\asretox, 0.1)$ } improves F1-Score over \etox both in devtest and test from 0.19 to 0.38. {We observe variations in recall rates for several HP languages (e.g., cat, heb, ind, rus, swh) between development and test datasets. This is due to the fact that we have very low representation of toxicity in devtest and test for those languages. That is why, it can be expected to have different results.}

\paragraph{MuTox Configurations} Supervised MuTox slightly improves zero-shot MuTox by 2\% on average. When comparing MuTox and ASR-MuTox averaging over 30 languages, results are almost comparable for zero-shot, but ASR-Mutox is better than MuTox in supervised setting. While it is unclear why MuTox vs. ASR-MuTox show different results depending on the language, we could hypothesize that the imbalances in the complexities of pronunciation/writing in various languages lead to variations in transcription quality. 

\begin{table*}[ht!]
\centering
\small
\begin{tabular}{lllllllllllllll}
\toprule
\textbf{lang}& \multicolumn{2}{c}{\textbf{Devtest (D)}} & \multicolumn{2}{c}{\textbf{Test (T)}} & \multicolumn{2}{c}{\textbf{MuTox$_{ZS}$}} & \multicolumn{2}{c}{\textbf{ASR-MuTox$_{ZS}$}} &  \multicolumn{2}{c}{\textbf{MuTox}} & \multicolumn{2}{c}{\textbf{ASR-MuTox}} & \multicolumn{2}{c}{\textbf{\detoxify}}\\ \midrule
 &  Size & Tox &  Size & Tox  & D & T & D & T &  D & T & D & T & D & T\\ \midrule
eng&1945&325&2918&486&-&-&-&-&0.61&0.63&0.72&0.75&0.68&0.71 \\
spa&1960&390&2941&585&-&-&-&-&0.63&0.65&0.72&0.73&0.69&0.71 \\\midrule
arb&720&63&1440&125&0.73&0.77&0.76&0.79&0.84&0.82&0.81&0.83&-&- \\
ben&738&20&1474&38&0.88&0.87&0.88&0.81&0.84&0.88&0.90&0.89&-&- \\
bul&675&95&1350&191&0.73&0.71&0.75&0.76&0.80&0.79&0.81&0.78&-&- \\
cat&606&22&1213&43&0.80&0.71&0.80&0.77&0.77&0.69&0.79&0.75&-&- \\
ces&744&43&1488&85&0.72&0.85&0.75&0.85&0.73&0.85&0.75&0.86&-&- \\
cmn&744&70&1487&140&0.78&0.79&0.76&0.77&0.82&0.78&0.80&0.74&-&- \\
dan&737&133&1481&266&0.82&0.82&0.86&0.83&0.83&0.82&0.90&0.85&-&- \\
deu&744&182&1486&362&0.77&0.80&0.78&0.79&0.83&0.85&0.85&0.86&-&- \\
ell&737&59&1474&118&0.71&0.70&0.79&0.78&0.71&0.68&0.80&0.79&-&- \\
est&684&86&1369&173&0.79&0.77&0.81&0.78&0.87&0.85&0.84&0.83&-&- \\
fas&743&40&1487&81&0.85&0.83&0.79&0.81&0.84&0.86&0.84&0.84&-&- \\
fin&738&87&1476&173&0.89&0.88&0.92&0.90&0.90&0.91&0.93&0.93&-&- \\
fra&738&62&1477&124&0.80&0.80&0.76&0.78&0.81&0.80&0.77&0.80&0.79&0.83 \\
heb&725&30&1450&60&0.73&0.71&0.72&0.76&0.70&0.74&0.71&0.75&-&- \\
hin&699&82&1399&166&0.77&0.77&0.77&0.79&0.80&0.83&0.81&0.84&-&- \\
hun&749&88&1499&177&0.72&0.77&0.77&0.80&0.77&0.76&0.76&0.81&-&- \\
ind&745&72&1490&143&0.70&0.73&0.67&0.69&0.74&0.76&0.73&0.75&-&- \\
ita&693&98&1385&197&0.61&0.60&0.63&0.60&0.64&0.66&0.62&0.69&0.79&0.62 \\
nld&729&87&1458&174&0.82&0.82&0.70&0.72&0.86&0.87&0.77&0.76&-&- \\
pol&741&64&1484&129&0.87&0.84&0.88&0.86&0.84&0.86&0.92&0.88&-&- \\
por&724&109&1449&218&0.78&0.76&0.80&0.78&0.79&0.75&0.78&0.76&0.81&0.83 \\
rus&741&81&1481&161&0.79&0.76&0.80&0.82&0.81&0.75&0.83&0.83&0.84&0.81 \\
slk&741&45&1482&90&0.82&0.84&0.84&0.84&0.81&0.87&0.83&0.87&-&- \\
swh&729&45&1458&89&0.69&0.69&0.66&0.66&0.70&0.68&0.69&0.67&-&- \\
tgl&736&43&1473&88&0.74&0.73&0.69&0.70&0.77&0.77&0.76&0.74&-&- \\
tur&740&53&1480&107&0.80&0.74&0.78&0.73&0.82&0.78&0.79&0.81&0.80&0.82 \\
urd&741&139&1486&278&0.78&0.79&0.76&0.78&0.81&0.83&0.79&0.84&-&- \\
vie&738&93&1477&185&0.78&0.81&0.75&0.76&0.81&0.84&0.81&0.80&-&- \\\midrule
\textbf{avg7}&-&-&-&-&0.75&0.73&0.75&0.74&0.73&0.72&0.75&\textbf{0.77}&\textbf{0.77}&0.76 \\
\textbf{avg}&-&-&-&-&0.77&0.77&0.77&0.78&0.78&0.79&\textbf{0.79}&\textbf{0.80}&-&- \\
\bottomrule
\end{tabular}
\caption{Toxicity detection AUC results of MuTox vs \detoxify. We show different MuTox configurations: ZS, trained only with English and Spanish; supervised, trained on English, Spanish, and HP languages; in audio speech (MuTox) or text (ASR-MuTox). The best results are bolded.\label{tab:toxicity:aucresults}}
\end{table*}

\begin{table*}[ht!]
\centering
\small
\begin{tabular}{lllllllllllllll}
\toprule
\multirow{2}{*}{\bf lang} & \multicolumn{4}{c}{\textbf{\etox}} & \multicolumn{2}{c}{\textbf{MuTox$_{ZS}$}} & \multicolumn{2}{c}{\textbf{ASR-MuTox$_{ZS}$}} &    \multicolumn{2}{c}{\textbf{MuTox}} & \multicolumn{2}{c}{\textbf{ASR-MuTox}}\\ \cmidrule(r){2-5}\cmidrule(lr){6-7}\cmidrule(l){8-9}\cmidrule(lr){10-11}\cmidrule(l){12-13}
& \multicolumn{2}{c}{Devtest} & \multicolumn{2}{c}{Test} & Devtest & Test & Devtest & Test  & Devtest & Test & Devtest & Test\\
&Prec & Rec & Prec & Rec & \multicolumn{8}{c}{Recall (Rec)}\\
\midrule
eng&0.39&0.30&0.40&0.31&-&-&-&-&0.18&0.23&0.58&0.67\\
spa&0.40&0.33&0.41&0.33&-&-&-&-&0.19&0.32&0.73&0.74\\\midrule
arb&0.09&0.02&0.21&0.04&0.16&0.26&0.17&0.15&0.60&0.54&0.44&0.53\\
ben&0.02&0.05&0.00&0.00&0.35&0.11&0.10&0.03&0.30&0.45&0.45&0.50\\
bul&0.19&0.29&0.17&0.21&0.22&0.32&0.53&0.65&0.71&0.69&0.74&0.65\\
cat&0.09&0.36&0.08&0.30&0.18&0.07&0.18&0.19&0.59&0.05&0.77&0.16\\
ces&0.12&0.44&0.18&0.73&0.19&0.45&0.30&0.59&0.21&0.41&0.30&0.61\\
cmn&0.00&0.00&0.00&0.00&0.14&0.37&0.27&0.17&0.43&0.35&0.41&0.16\\
dan&0.26&0.77&0.23&0.68&0.92&0.91&0.93&0.90&0.89&0.92&0.96&0.88\\
deu&0.42&0.36&0.43&0.36&0.81&0.88&0.95&0.99&0.86&0.90&0.98&0.99\\
ell&0.11&0.39&0.12&0.41&0.02&0.16&0.03&0.36&0.19&0.18&0.44&0.31\\
est&0.15&0.49&0.15&0.51&0.57&0.54&0.71&0.63&0.86&0.84&0.80&0.76\\
fas&0.07&0.75&0.07&0.68&0.15&0.23&0.18&0.33&0.33&0.14&0.50&0.36\\
fin&0.17&0.90&0.17&0.89&0.91&0.86&0.94&0.94&0.87&0.88&0.94&0.94\\
fra&0.11&0.42&0.10&0.39&0.26&0.31&0.08&0.14&0.40&0.53&0.48&0.50\\
heb&0.05&0.63&0.04&0.53&0.10&0.05&0.07&0.10&0.13&0.05&0.03&0.70\\
hin&0.18&0.02&0.06&0.01&0.56&0.36&0.54&0.33&0.63&0.72&0.66&0.79\\
hun&0.20&0.73&0.22&0.79&0.34&0.51&0.57&0.57&0.55&0.49&0.56&0.64\\
ind&0.10&0.22&0.11&0.28&0.18&0.19&0.07&0.17&0.15&0.41&0.32&0.40\\
ita&0.20&0.38&0.16&0.27&0.10&0.05&0.28&0.18&0.16&0.21&0.31&0.35\\
nld&0.08&0.15&0.06&0.11&0.77&0.73&0.21&0.21&0.78&0.84&0.61&0.52\\
pol&0.12&0.89&0.12&0.88&0.86&0.60&0.83&0.73&0.67&0.73&0.86&0.79\\
por&0.28&0.43&0.30&0.50&0.68&0.69&0.75&0.77&0.74&0.59&0.66&0.63\\
rus&0.17&0.44&0.19&0.47&0.54&0.42&0.47&0.52&0.65&0.27&0.69&0.63\\
slk&0.14&0.73&0.09&0.48&0.49&0.41&0.36&0.52&0.47&0.56&0.49&0.66\\
swh&0.08&0.18&0.07&0.15&0.04&0.02&0.64&0.06&0.02&0.61&0.60&0.46\\
tgl&0.27&0.09&0.12&0.03&0.19&0.05&0.05&0.06&0.35&0.23&0.07&0.09\\
tur&0.06&0.08&0.12&0.16&0.13&0.93&0.15&0.12&0.34&0.28&0.28&0.47\\
urd&0.00&0.00&0.00&0.00&0.87&0.87&0.81&0.85&0.88&0.88&0.83&0.92\\
vie&0.12&0.16&0.10&0.15&0.54&0.69&0.46&0.61&0.68&0.80&0.73&0.68\\
\midrule
\textbf{avg} &0.17 &	0.36&	0.17&	0.35&	0.40 &	0.43&	0.42&	0.42 &	0.47&	0.48&	\textbf{0.57}&	\textbf{0.58} \\
\bottomrule
\end{tabular}
\caption{Toxicity detection precision and recall results. MuTox recall at the precision of $max (\etox, 0.3)$ vs. \etox. The best results are bolded.\label{tab:toxicity:precrecallresults}}
\end{table*}


\paragraph{Performance across toxicity categories} Figure \ref{fig:categories2} reports the model performances by toxicity categories. We note that \detoxify and MuTox have less variance (<0.001) across categories than \etox (>0.01).
\etox performs well on Profanities (recall>0.8), since those can easily be identified with explicit wordlists. It struggles however with more complex/implicit types of toxicity such as Hate Speech or Physical Violence.


\begin{figure}[ht!]
\center
    \includegraphics[width=7.9cm]{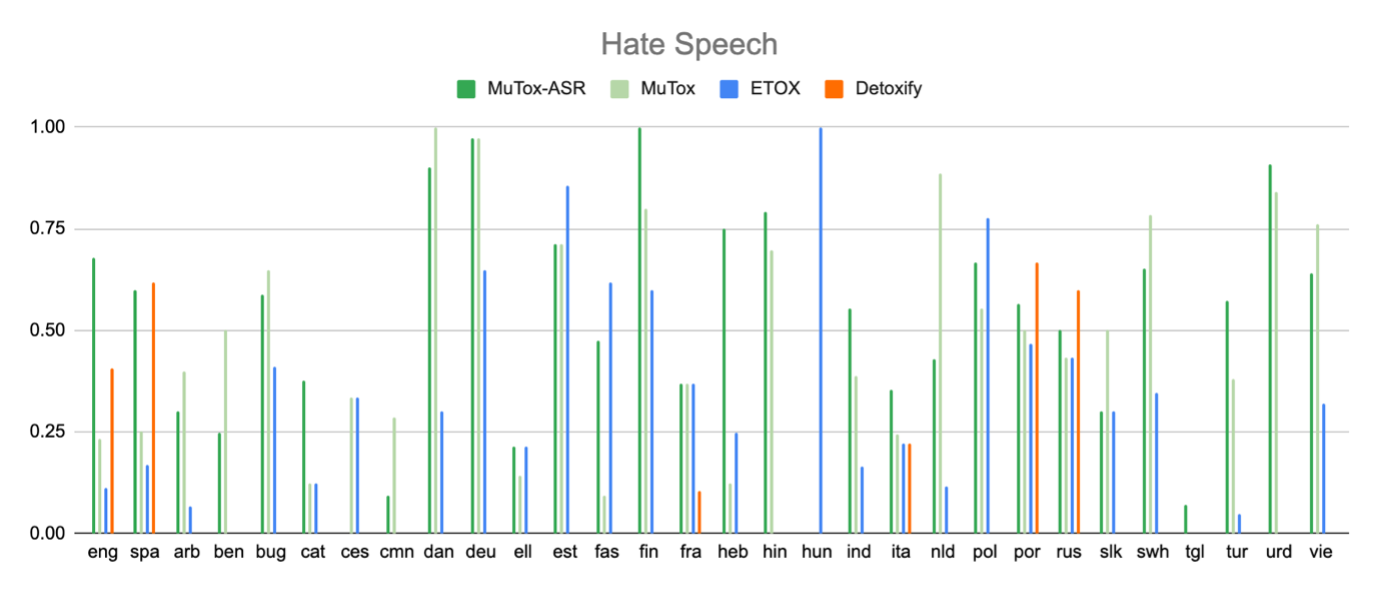}
    \includegraphics[width=7.9cm]{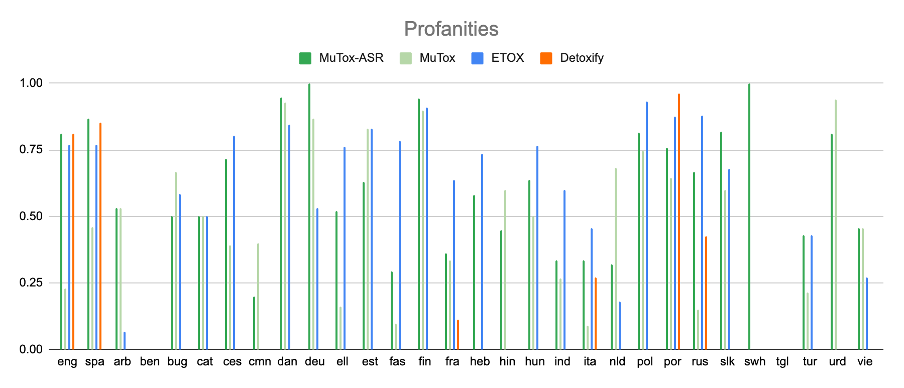}
    \includegraphics[width=7.9cm]{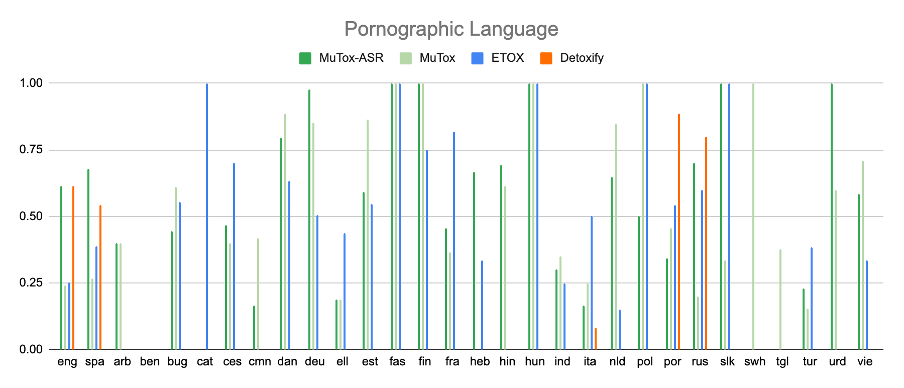}
    \includegraphics[width=7.9cm]{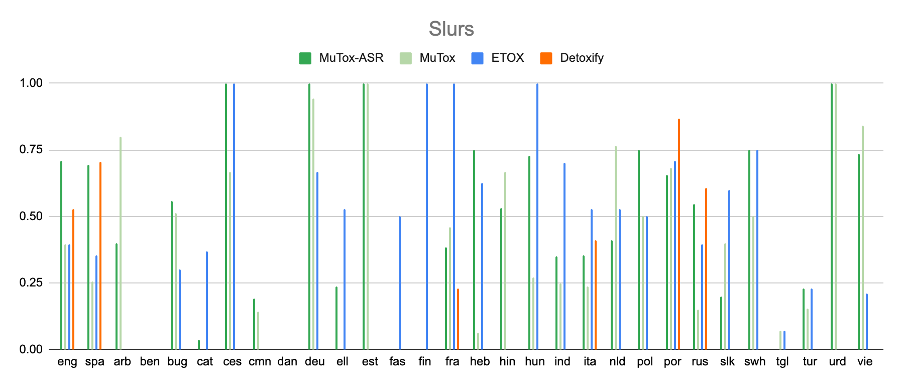}
    \includegraphics[width=7.9cm]{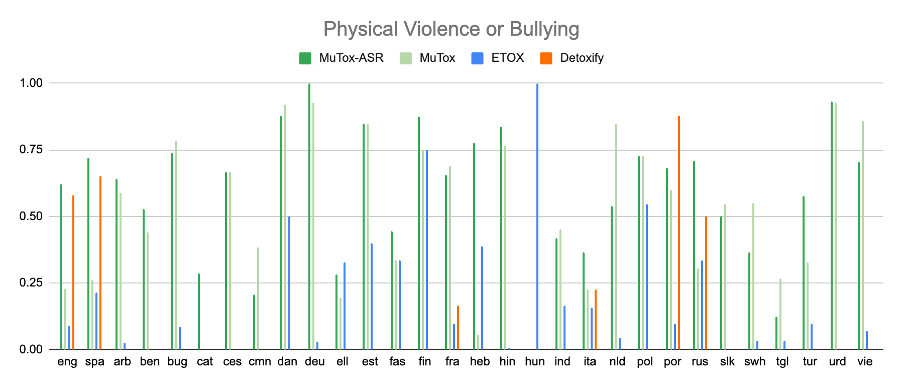}
    \caption{Recall per toxicity category at fixed precision of  $max(\etox, 0.3)$.}    \label{fig:categories2}
\end{figure}






\section{Conclusions}
\label{sec:conclusion}

In this paper, we introduced MuTox, a highly multilingual dataset and a massive multilingual toxicity detector that lays the groundwork for the largely unexplored task of multilingual audio-based toxicity detection.
Our MuTox dataset enables benchmarking of multilingual audio-based toxicity detection across 30 languages. Our MuTox classifier, compared to cascade tools (speech recognition followed by text-based toxicity classifiers) that have similar coverage (more than hundreds of languages), shows superior performance for all evaluated tasks.

Data\footnote{Will be revealed after the anonymized review.
} and code\footnote{Will be revealed upon anonymized review.
} with their corresponding data and model cards are freely available for further research and development. 
In future work, {we primarily aim to explore more complex architectures for training MuTox detector. Additionally, we want to evaluate MuTox's performance on the task of added toxicity. We also intend to use it to analyze and improve ETOX wordlists, as detailed in the appendix \ref{app:wordlists}. This ongoing work will continue to advance the field of multilingual audio-based toxicity detection.}


\section*{Ethical Considerations and Limitations}

\paragraph{Annotators.} Annotations were provided by professional annotators, who were informed of the nature of the content to be annotated and were given the opportunity to opt out. {We understand that the perception of toxicity is subjective, and therefore varies greatly from individual to individual, and from group of individuals to group of individuals. Even within groups of any size, the probability is high that toxicity will be perceived differently from one individual to the next. However, we involved a large pool of annotators with native level in each of the MuTox languages. While our guidelines specify the types of toxicity annotators should look out for, it is ultimately down to annotator judgment to interpret whether a term or phrase is toxic, given the context, and given their personal interpretation. }

\paragraph{Bias in pre-selected sentences.} Our dataset may be biased towards text-based toxicity because we pre-selected annotations using this criterion. The use of different text classifiers, as well as a variety of thresholds, mitigates this potential bias. {Moreover, the data is sourced from (e.g. \commonvoice), and it represents a broad and geographically diverse sample. }

\paragraph{Unintended bias.} Our evaluation does not cover unintended biases\footnote{\url{https://www.kaggle.com/competitions/jigsaw-unintended-bias-in-toxicity-classification/}}, which we intend to cover in future work.

\section*{Acknowledgements}

The authors want to thank Samuel Bell for enriching discussions and his valuable feedback.

\bibliography{anthology,custom}
\bibliographystyle{acl_natbib}

\appendix

\section{Languages}
\label{app:lang}

\begin{table}[h!]
\centering
\small
\begin{tabular}{lll}
\toprule
arb&	Modern Standard Arabic & Mesopotamian Arabic\\
ben&	Bengali & Indo-Aryan\\
bul&	Bulgarian & Balto-Slavic\\
cat&	Catalan & Romance\\
ces&	Czech & Balto-Slavic\\
cmn&	Mandarin Chinese  & Sino-Tibetan\\
dan&	Danish & Germanic\\
deu&	German & Germanic \\
ell&	Greek & Hellenic\\
eng&	English & Germanic\\
est&	Estonian & Uralic \\
fas&	Western Persian & Iranian\\
fin&	Finnish & Uralic\\
fra&	French & Romance\\
heb & Hebrew & Afro-Asiatic\\
hin&	Hindi & Indo-Aryan\\
hun&	Hungarian & Uralic\\
ind&	Indonesian &Austronesian\\
ita&	Italian & Romance\\
nld & Dutch & Germanic\\
pol&	Polish & Romance\\
por&	Portuguese & Romance\\
rus&	Russian & Balto-Slavic\\
spa&	Spanish & Romance\\
slk&	Slovak & Balto-Slavic\\
swh&	Swahili & Atlantic-Congo\\
tgl&	Tagalog & Austronesian\\
tur&	Turkish & Turkic\\
urd&	Urdu & Indo-Aryan\\
vie&	Vietnamese & Austroasiatic\\
\bottomrule
\end{tabular}
\caption{The 30 languages covered in this work.}
\label{table:language_list}
\end{table}

\section{Annotations Details}
\label{app:quantiles}

Figure \ref{fig:quantiles} shows the percentage of toxicity obtained with the annotation (y-axis) by the quantile of toxicity (x-axis) in the text data. 

\begin{figure}[ht!]
\center
    \includegraphics[width=7.9cm]{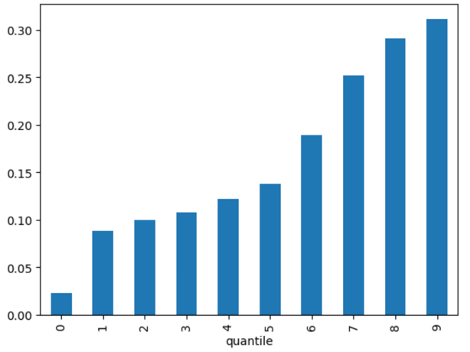}
    \includegraphics[width=7.9cm]{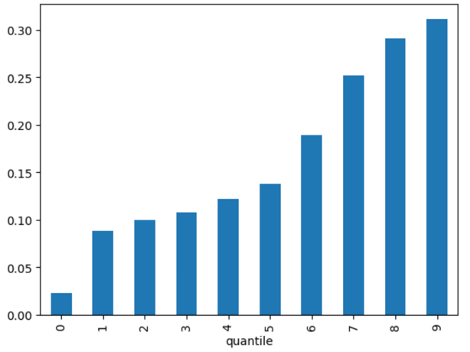}
    \caption{Percentage (y-axis) of toxicity in the audio speech dataset in English (top) and in Spanish (bottom) per toxicity quantile (x-axis) in the text toxicity classification.}
    \label{fig:quantiles}
\end{figure}

\section{Wordlists analysis.} 
\label{app:wordlists}

We report an analysis on the toxic words detected with \etox in order to understand the limitations of the Toxicity-200 word-lists.

For English we have a total of 110 different toxic tokens detected and 59 tokens show reasonable precision (>0.4). On the one hand, the worst performing tokens are insults such as stupid* and fool*, these insults are not so harsh and probably some times are not considered non-toxic by native speakers. These show a high output, but very low precision and hence are negatively affecting \etox overall's performance. On the other hand, some of the best performing tokens are variations of fuck* and shit*. These tokens show a high output and precision (>0.8), being responsable for a considerable share of recall (>0.15). Slurs tend to have a high precision but low recall.

For Spanish we have a total of 187 different toxic tokens detected and 110 tokens show reasonable precision (>0.4). The worst performing tokens are insults such as tonto* and maldito* and some of the best performing tokens are  are m*erda and variations of j*der. The recall is much more distributed compared to English, with several terms having a precision above 0.4. 

\begin{figure}[ht!]
\center
    \includegraphics[width=7.9cm]{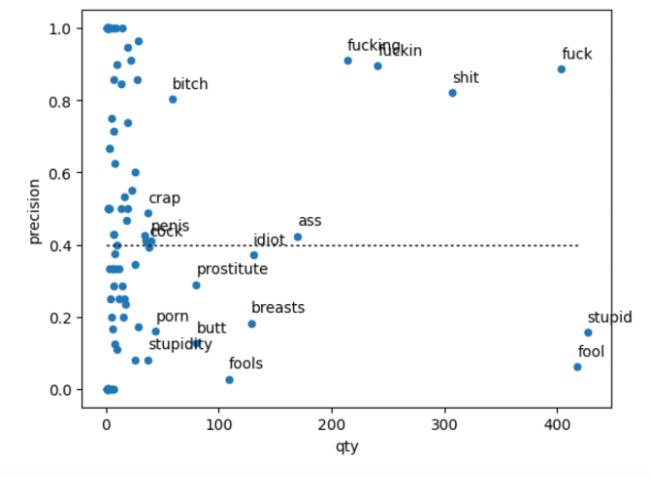}
        \includegraphics[width=7.9cm]{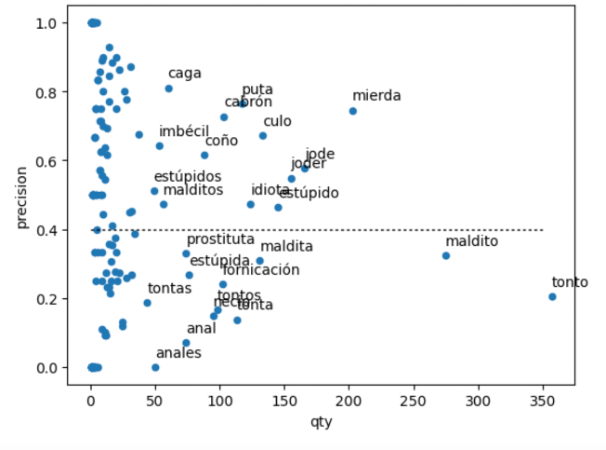}
    \caption{ETOX's tokens performance for our English Dataset (top) and Spanish Dataset (bottom). Vertical axis representing the precision and Horizontal axis the total output.}
    \label{fig:wordlist}
\end{figure}

This analysis could be further extended to HP languages.

\end{document}